# RELIABILITY MODELS FOR SMARTPHONE APPLICATIONS

Sonia Meskini [1], Ali Bou Nassif [2] and Luiz Fernando Capretz [3][0000-0001-6966-2369]

[1] *Department of Electrical & Computer Engineering, Western University, Canada*
*sonya.meskini@gmail.com*
[2] *Department of Electrical & Computer Engineering, University of Sharjah, UAE*
*anassif@sharjah.ac.ae*
[3] *Department of Electrical & Computer Engineering, Western University, Canada*
*lcapretz@uwo.ca*

**Abstract.** Smartphones have become the most used electronic devices. They carry out most of the functionalities of desktops, offering various useful applications that suit the user's needs. Therefore, instead of the operator, the user has been the main controller of the device and its applications, therefore its reliability has become an emergent requirement. As a first step, based on collected smartphone applications failure data, we investigated and evaluated the efficacy of Software Reliability Growth Models (SRGMs) when applied to these smartphone data in order to check whether they achieve the same accuracy as in the desktop/laptop area. None of the selected models were able to account for the smartphone data satisfactorily. Their failure is traced back to: (i) the hardware and software differences between desktops and smartphones, (ii) the specific features of mobile applications compared to desktop applications, and (iii) the different operational conditions and usage profiles. Thus, a reliability model suited to smartphone applications is still needed. In the second step, we applied the Weibull and Gamma distributions, and their two particular cases, Rayleigh and S-Shaped, to model the smartphone failure data sorted by application version number and grouped into different time periods. An estimation of the expected number of defects in each application version was obtained. The performances of the distributions were then compared amongst each other. We found that both Weibull and Gamma distributions can fit the failure data of mobile applications, although the Gamma distribution is frequently more suited.

## 1. Introduction

Smartphones are now so useful that many people prefer them over desktop or laptop computers. These mobile devices are overselling desktop and laptop computers. They are becoming a necessary commodity to many users and their prices have been decreasing. Hundreds of applications, usually suited to desktop or laptop computers, have been adapted to and carried out by these smartphones. Owing to their small size, other specific applications have been built in, ranging from simple applications to very critical ones. The high usage and trust placed in these devices and their applications make their reliability a critically important goal to achieve [1]. Thus, manufacturers are competing to release the most reliable devices, and they have successfully achieved high reliability, in terms of hardware, by applying traditional and enhanced Hardware Reliability Growth Models (HRGM) [2]. These HRGMs have been useful for classic mobile phones. On the other hand, owing to their highly integrated software, smartphones are far more advanced devices and their functionalities far exceed those of the classic mobile phones. Therefore, increased attention is now being paid to the reliability and security of these devices.

Many companies in the mobile business, as they expand rapidly under pressure from the market and competition, do not use appropriate software engineering methods to develop their products and services [3, 4]. As a result, their software is less reliable and even more expensive than it should be. Therefore, the reliability issue is becoming as acute in the mobile area [5, 6] as for desktop and laptop computers. Furthermore, owing to the peculiarities of the Development Life Cycle (DLC) of mobile application software, the reliability issues [7] in the mobile area are likely to differ from those in the desktop or laptop area.

Software Reliability Growth Models (SRGMs) are among the tools that deal with the reliability of software applications; they have been constructed and successfully applied to desktop (classic/standard) applications. In recent work [8], we thoroughly investigated the applicability of these SRGMs to the mobile area. We applied three of the most used SRGMs to the collected failure data of three smartphone applications; our main conclusion was that none of the selected models was able to account for the observed failure data satisfactorily. Their failure was traced back to the specific features of mobile applications compared to desktop applications. Pursuing our investigation, and after collecting further failure data from many versions of the same applications, we addressed the following research questions:

(1) How do the existing successful reliability models, used to assess the desktop/laptop applications, perform when applied to the mobile area?
(2) What are the best non-linear distributions that fit smartphone application failure data?
(3) What useful information can be gained from this approach?

The rest of the paper is organized as follows: in Section 2 we present a brief comparative study of smartphones versus desktop or laptop computers in order to stress the main differences and their incidence for the reliability of mobile applications. In Section 3 we provide a short roadmap of the existing models that we will use, we describe our dataset collection, and we test the applicability of existing software reliability models. Finally, in Section 4, we carry out an analysis of the failure data with model distributions followed by a discussion in Section 5 and the threats to validity in Section 6. We present our conclusions in Section 7 and outline future work possibilities.

## 2. Reliability for Smartphone Applications

Smartphones have become personal devices that are used almost anywhere, at any time, and for practically everything. High expectations and trust are placed on this mobile device, and it is used as more than just a phone and the usage exceeds the functionalities of only sending SMS or making voice calls.

A few years ago, smartphone usage was limited to business. Lately, thanks to network and mobile technology improvements and progress, those smart devices have begun to gain traction and they have received remarkable acceptance in the user market. Since then there has been a significant increase in smartphone technology. Countless smartphone applications have been designed and developed. Along with this, the market has been growing rapidly, and market analysis has confirmed that it will continue growing to the point that it will exceed growth in desktop or laptop computers and oversell many other electronic devices, in particular laptop computers.

As a result, various types of customer needs must be satisfied to ensure continuing growth in this competitive market because previous studies have shown that those needs differ from one region to another and from one user to another [1]. In this situation the reliability of smartphone applications will play an important role to keep user trust in their devices.

Nevertheless, the quality and achievement levels of smartphone applications are not keeping pace with the increase in number of applications being developed. This is a consequence of the “time to market” strategy that the companies developing smartphone applications have adopted. For developers to meet project deadlines in designing these new applications, they have overlooked some development phases (especially the design phase, which is considered as the most important stage) of the DLC of the application [9]. During the design phase, multiple defects and bugs might be avoided. Hence, following a strategy of overlooking phases could result in many problems later on that might take more time and effort to solve than was used to design the entire application.

In step with this, for companies to be competitive it is important to study the market using surveys and other analysis and to understand that location is one of the important factors in the variety of user requirements. Therefore, a competitive smartphone application must meet these requirements before, during and after the DLC.

However, anticipating and meeting some of those requirements might be a challenge for developers in the design stage; the resulting failures might be difficult to solve in the execution stage since identifying and fixing the cause of the failure can be more difficult at that stage. The reason for this is that there are many factors that could result in the unreliability of the application or its failure – these include the nature of the technology used, the platform, the version of the Operating System (OS), and many other internal or external factors [9, 10, 11].

In the following discussion, we will provide a short road map of the most famous SRGMs that have been successfully applied in desktop and laptop applications, and we will check whether they will achieve the same success in the mobile area.

## 3. SRGMs Applied to Smartphone Applications

This section is devoted to a presentation of the application of three SRGMs that are known to be successful in both the desktop and laptop area to three concrete cases of smartphone applications.

The SRGMs used later in our experiments are: the NHPP – Crow – AMSAA model (also termed the NHPP-Power Law model), the Musa-Basic execution time model (or the exponential model), and the Musa-Okumoto model (or the Logarithmic Poisson model).

The applications that have been chosen are Skype, Vtok, and a private Windows phone application.

- For the NHPP model the mean value function is:

$$\boldsymbol{\mu}(\boldsymbol{t}) = \boldsymbol{N}((1 - \mathrm{e}^{-\mathrm{bt}}); \boldsymbol{where}\ \{\boldsymbol{b}, \boldsymbol{N}\} > 0. \quad (1)$$

  N: total number of faults and b: fault removal rate.
- The Musa-Basic model, also termed the exponential model, is given by the following mean value:

$$\boldsymbol{\mu}\ (\boldsymbol{t}) = \boldsymbol{\beta}_{\mathbf{0}}^{\boldsymbol{E}}(\mathbf{1} - \boldsymbol{e}^{-\boldsymbol{\beta}_{\mathbf{1}}^{\boldsymbol{E}}\boldsymbol{t}}). \quad (2)$$

  where $\beta_0^E$ is the expected number of failures, and $\beta_1^E$ is the hazard rate.
- The Musa-Okumoto model, also termed the logarithmic model, is given by the following mean value:

$$\boldsymbol{\mu}\ (\boldsymbol{t}) = \boldsymbol{\beta}_{\mathbf{0}}^{\boldsymbol{L}}\boldsymbol{Ln}(\mathbf{1} + \boldsymbol{\beta}_{\mathbf{1}}^{\boldsymbol{L}}\boldsymbol{t}). \quad (3)$$

  where $\beta_0^L$ is the expected number of failures, and $\beta_1^L$ is the hazard rate.

The reasons for our choice of these SRGMs are: (1) these are based on a few simple and reasonable assumptions, (2) they are simple to understand on physical grounds, and (3) they are implemented in reliability tools like RGA7 and SMERFS.

We present the procedure devised to collect the failure data for each application followed by the results of the application of the chosen SRGM to failure data for each application, and, finally, an analysis of the observed results.

### 3.1. *Existing Reliability Models*

As a matter of fact, hundreds of models exist but those models were chosen because, according to a Software Reliability modeling survey [5, 12], they are the most useful and successful models in the computer applications domain. Hence, one of our research goals is to check if those models will achieve the same success when applied to smartphone applications failure data as was the case for the desktop and Hardware Growth Reliability Models [2].

The purpose behind developing models is to measure and estimate software reliability. This has become an important and major target for companies because reliability has a significant effect in each stage of the DLC of an application from the design to the maintenance stages [13, 14].

The first SRGM was developed in 1972 [5, 13]. An SRGM usually results in a set of mathematical equations that accurately fit the collected failure data [12]. Any model relies on simplifying assumptions; however, some of these assumptions may not be useful in real situations [12].

Due to the limited space in this paper we will not present all of the assumptions of the different models used later but only the basic ones that are shared by all of the models. The detailed list of the different assumptions of each model can be found in [5].

The common assumptions are: "*1) The software is operated in a similar manner as that in which reliability predictions are to be made, 2) Every fault has the same chance of being* *encountered within a severity class as any other fault in that class,* and *3) The failures, when the faults are detected, are independent"* [5].

### 3.2. *Datasets*

We used Apple devices (iPhone, iPad and iPod Touch) crash files as well as a Windows Phone crash file as our "experimental" data. These crash files are not public, but are confidential. Hence, we will focus more on the Apple devices crash files since it was easier to collect them through a survey that was sent to different people from different parts of the world. Many of those to whom we sent surveys agreed to send us their failure data, whereas other did not.

For the Windows Phone case, we could only get the crash file report of one application due to confidentiality policies. Collecting the data was, and still is, a challenge. Figure 1 presents an example of the Apple devices crash log.

1 1

```
Incident Identifier: 2D961565-D688-4CB4-A5EF-4F1BFF4620F9
CrashReporter Key:   4c67cf6e529b1b2ecc2c57df10d48b53f0bdbb50
Hardware Model:      iPhone4,1
Process:         Skype [3127]
Path:            /var/mobile/Applications/E3AF5F07-1C5A-4172-A40E-ACCA269519CB/Skype.app/Skype
Identifier:      Skype
Version:         ??? (???)
Code Type:       ARM (Native)
Parent Process:  launchd [1]

Date/Time:       2011-11-03 14:27:10.635 -0400
OS Version:      iPhone OS 5.0 (9A334)
Report Version:  104

Exception Type:  EXC_BAD_ACCESS (SIGSEGV)
Exception Codes: KERN_PROTECTION_FAILURE at 0x2fd00fe8
Crashed Thread:  0

Thread 0 name:  Dispatch queue: com.apple.main-thread
Thread 0 Crashed:
0   libsystem_c.dylib             0x380ca308 0x380be000 + 49928
1   CoreFoundation                0x3710d946 0x37071000 + 641350
2   CoreFoundation                0x3710cb9c 0x37071000 + 637852
```

**Fig. 1.** Apple crash file.

The crash log is a long text file full of symbols and information that we ordinarily do not need, yet it contains information that we used to create our failure dataset. To achieve that, we developed a program in JAVA that we run each time we synchronize the devices or receive log folders from other users to update our dataset.

Figure 2 shows an example of the output file of the JAVA program developed for the purpose of extraction; in this example *Identifier* is the name of the application, *Date/Time* is the date and time of the crash, and *Crashed Thread* is the number of the thread that caused the crash.

```
Identifier:      Skype
Date/Time:       2012-02-25 01:58:19.603 +0300
Crashed Thread:  0
Identifier:      Skype
Date/Time:       2012-02-26 00:15:58.353 +0300
Crashed Thread:  0
Identifier:      Skype
Date/Time:       2012-02-26 00:16:50.428 +0300
```

**Fig. 2.** Output of the JAVA program.

### 3.3. *Experiments*

The reliability demonstration of smartphone applications was carried out through traditional testing, failure data collection, and the application of the most used SRGMs for standard applications to observe and check the adequacy of these models in the mobile area.

For this purpose we used two applications for iOS and one for Windows mobile phone to test the models with different platforms. We could not collect enough data from Android phones but we are still collecting to eventually have enough data to test the models on Android applications.

The first iPhone application studied was Skype, which had been tested and used for one year (from November 1, 2011 to November 11, 2012). Hence, the data has been collected during this year with some missing values due to the occasional non-use of the application. We were, however, able to collect 39 data points for the Skype application.
The second application studied was Vtok (an application for Google talk). This application was used continuously every day for two months (from September 19, 2012 to November 25, 2012). Hence, we were able to collect failures every day (81 data points).

Each of the above mentioned SRGM models was applied to Skype and Vtok failure data, which represent two different situations: the Skype application used during one year but with some missing values, and the Vtok application used every day for two months, with the possibility of collecting more than one failure per day. This is an instance of testing the efficiency and accuracy of the models in different situations with different types of data.

During these periods, both the Skype and the Vtok applications were upgraded when new versions were released.

On the other hand, the Windows phone application was used and tested continuously for six months (from March 2012 to August 2012) by different users located in different parts of the world (more than 100 users). The crash count of the application is illustrated in Figure 3.

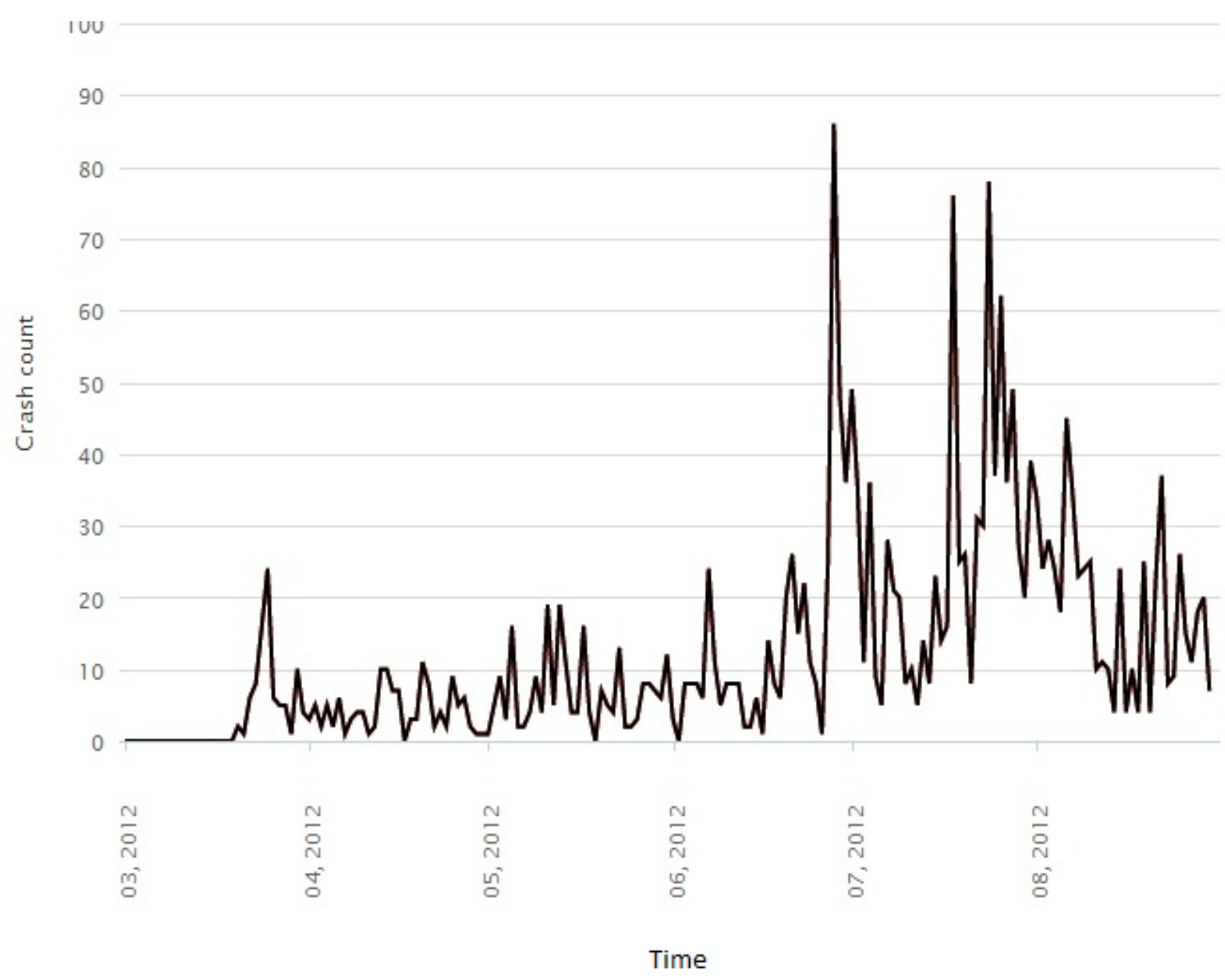

**Fig. 3.** Windows Phone crash count.

We noted that June, July, and August are the months with the highest crash rate. Since this application is developed for the purpose of locating bicycle stations, it receives more use during the summer than in winter, which explains the high crash rate during the hot season. This highlights the fact that the type of application and its usage play an important role in its reliability. From this graph we extracted failure data during six months (177 data points).

We used two Software Reliability tools for this application to double check the results. The first tool is RGA 7 from ReliaSoft and the second one is **S**tatistical **M**odeling and **E**stimation of **R**eliability **F**unctions for **S**oftware (SMERFS). We configured our tools as follows: we choose 1 for the severity level of all failures and one hour for the unit. As the time scales of the three applications are very different, we choose to normalize our data between 0 and 1.

As the RGA tool does not accept zero value as a time to event, we entered 0.001 instead of 0 as the first value to be able to have results. For the severity, 1 was selected because the applications used are not going to cause harmful consequences if they fail. But this is not the case with other applications. When working with applications such as online banking, health, and stock exchange, etc. the severity of the failure has to be taken into consideration.

### 3.4. *Evaluation*

Smartphone application reliability is a challenge. Thus, it is a necessity for reliability methods being used elsewhere to be evaluated and to assess their validity in the mobile area. One of the main goals of this work is to check if the most accurate and most used SRGMs for desktop applications have the same success and accuracy when applied to smartphone applications.

Figure 4 presents the cumulative number of failures per time for the Skype application when applying the NHPP model. The RGA tool indicates an evident failure.

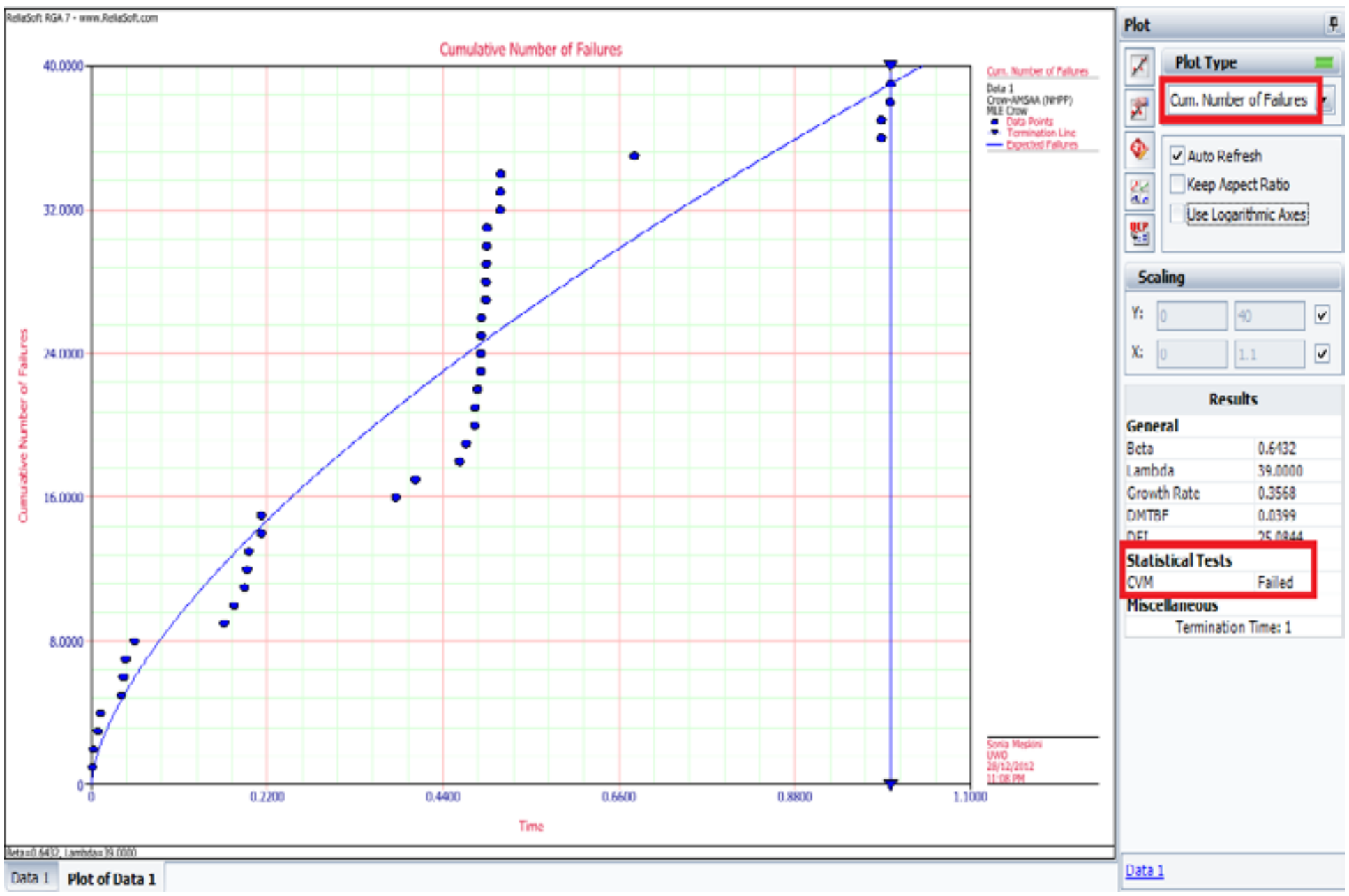

**Fig. 4.** Cumulative number of failures per Time (Skype).

Figure 5, represents the cumulative number of failures per time for the Vtok application. Again the NHPP model failed to fit the data.

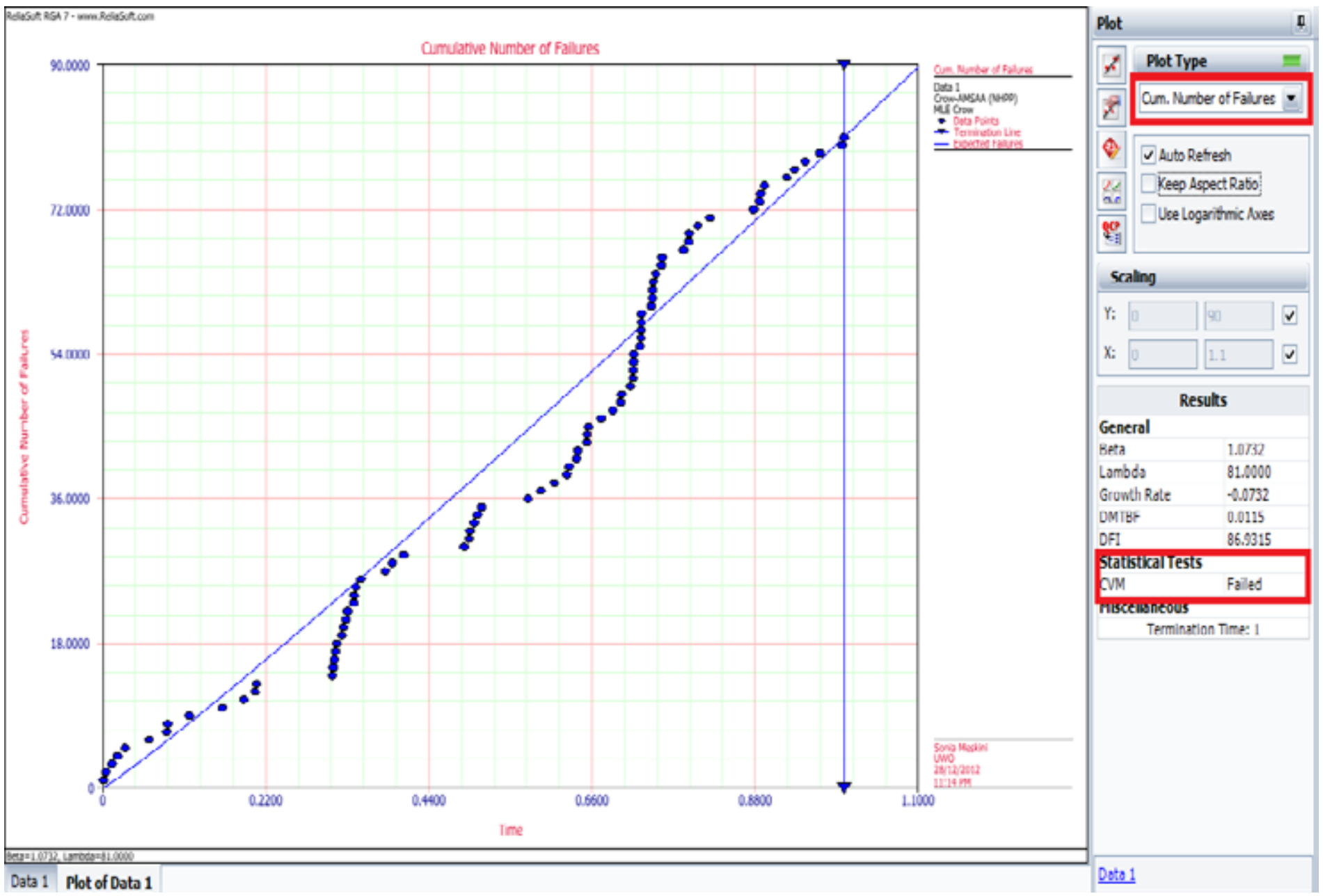

**Fig. 5.** Cumulative number of failures per Time (Vtok).

As mentioned in the previous section, we used Skype for a year and collected failure data that contains some missing values. Failure data for the Vtok application that was used for two months showed more than one failure per day. However, the NHPP still fails to fit these two different types of data. One reason is that failure data is a dynamic process for mobile applications, which means that the occurring number of failures is unpredictable, sometimes decreasing and sometimes increasing. For example, in Figure 4 from t = 0.2076 until t = 0.3097 the application did not experience a failure and from t = 0.3097 until t = 0.3484 an important number of failures occurred.

In order to confirm our results we used a second tool, SMERFS, and we applied the NHPP model to the same data points. The result was the same – the failure of the model each time.

Figure 6 and Figure 7, for Skype failure data, and Figure 8, for Vtok failure data, show the results when applying the Musa-Basic and Musa-Okumoto models. Each time the models failed to fit the data. In fact, the models failed completely to fit the Vtok failure data (Figure 8).

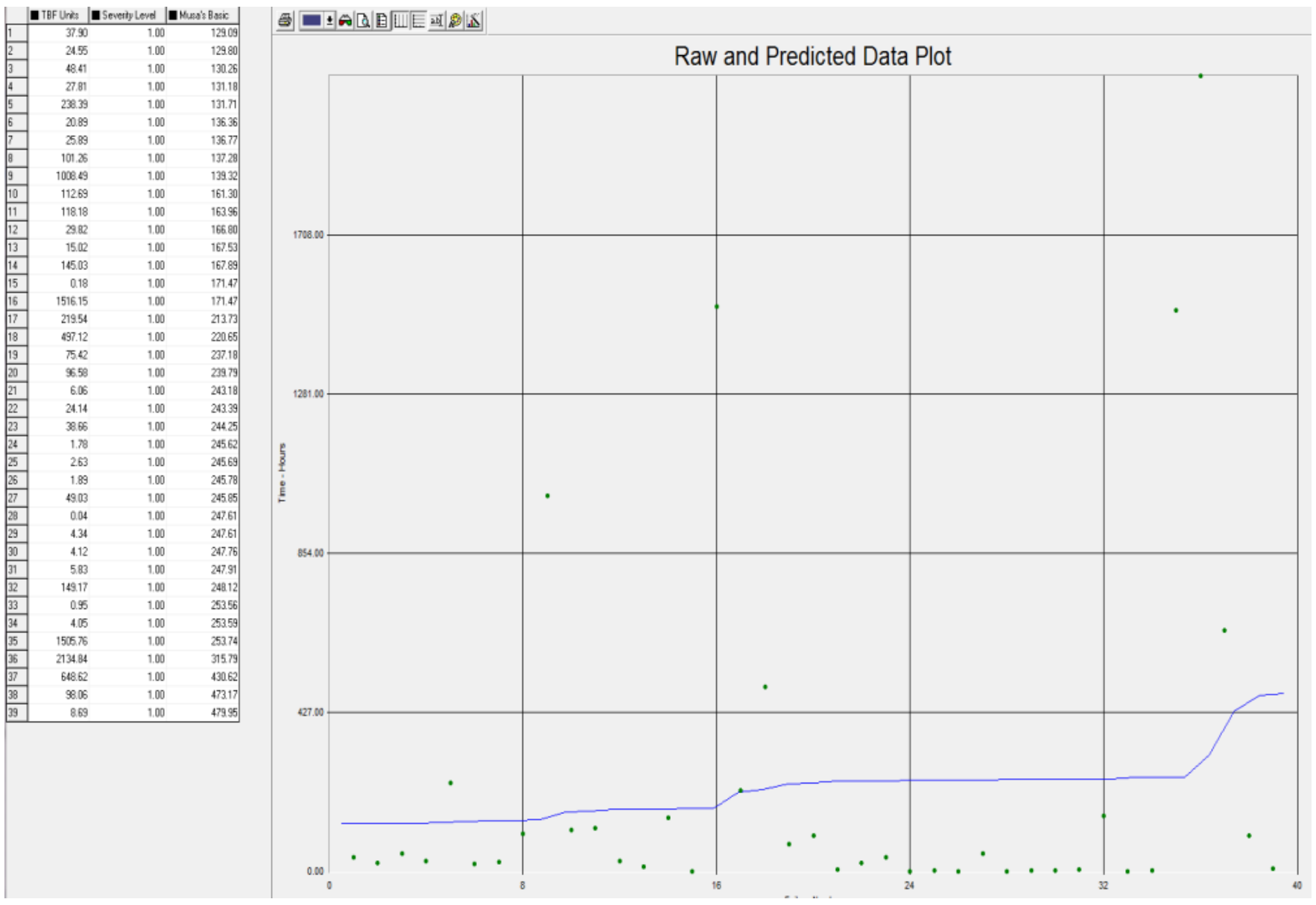

**Fig. 6**. Skype failure data and failure of the Musa-Basic model.

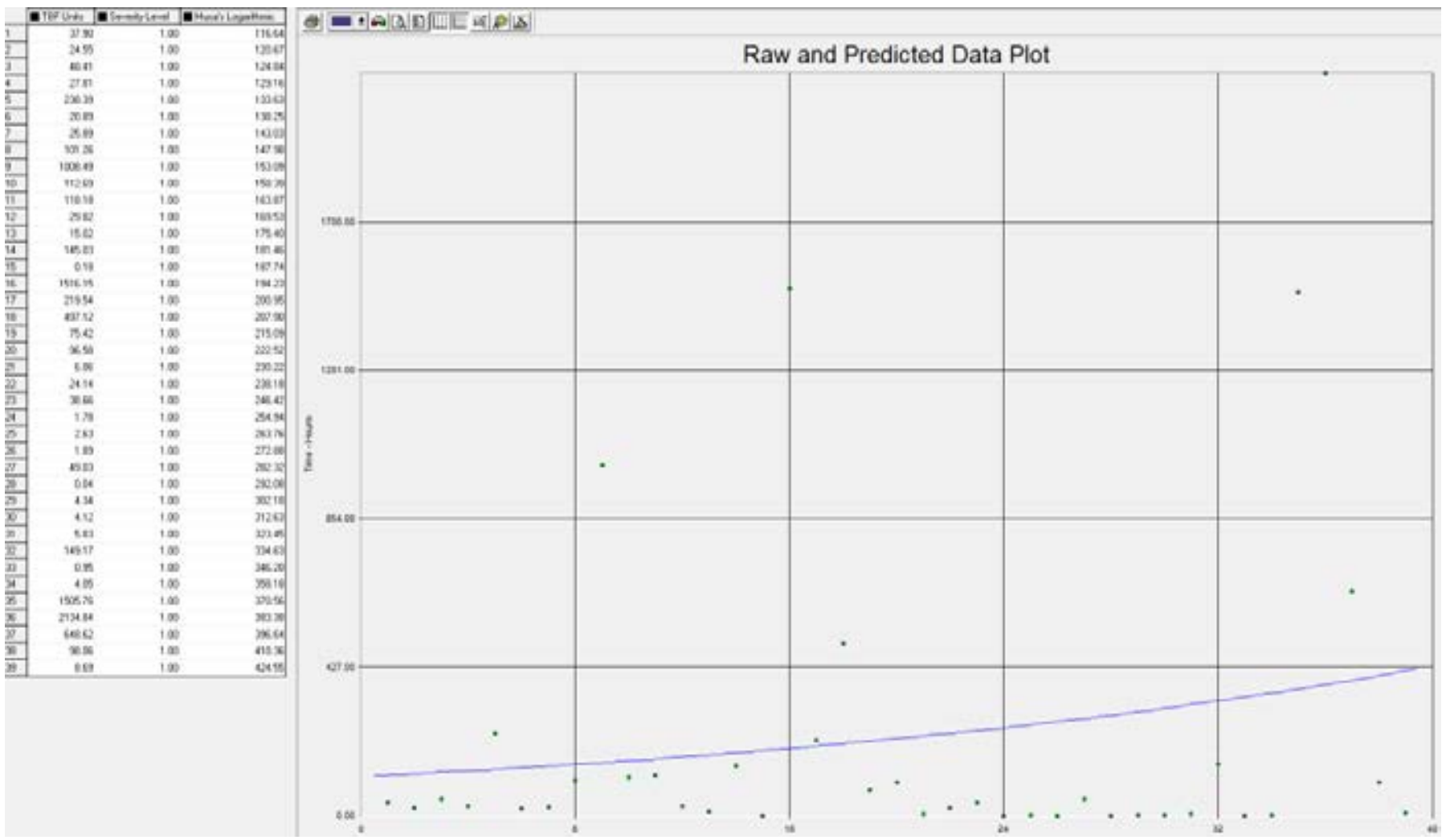

**Fig. 7.** Skype failure data and failure of the Musa-Okumoto model.

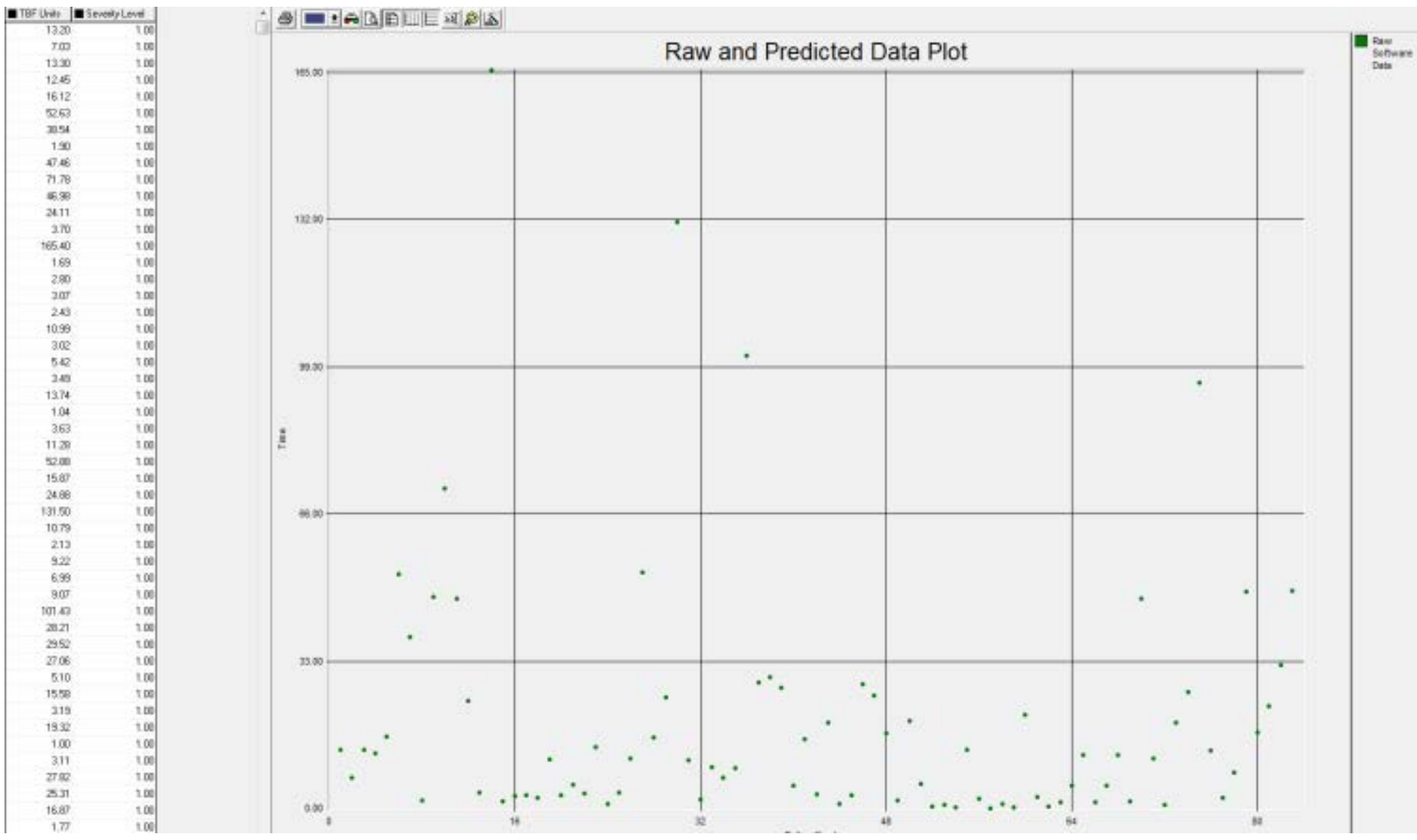

**Fig. 8.** Vtok failure data and failure of the three selected models.

Figure 9 represents the results of the application of the NHPP model to the Windows Phone failure data. Once again, the RGA tool indicates the failure of the model.

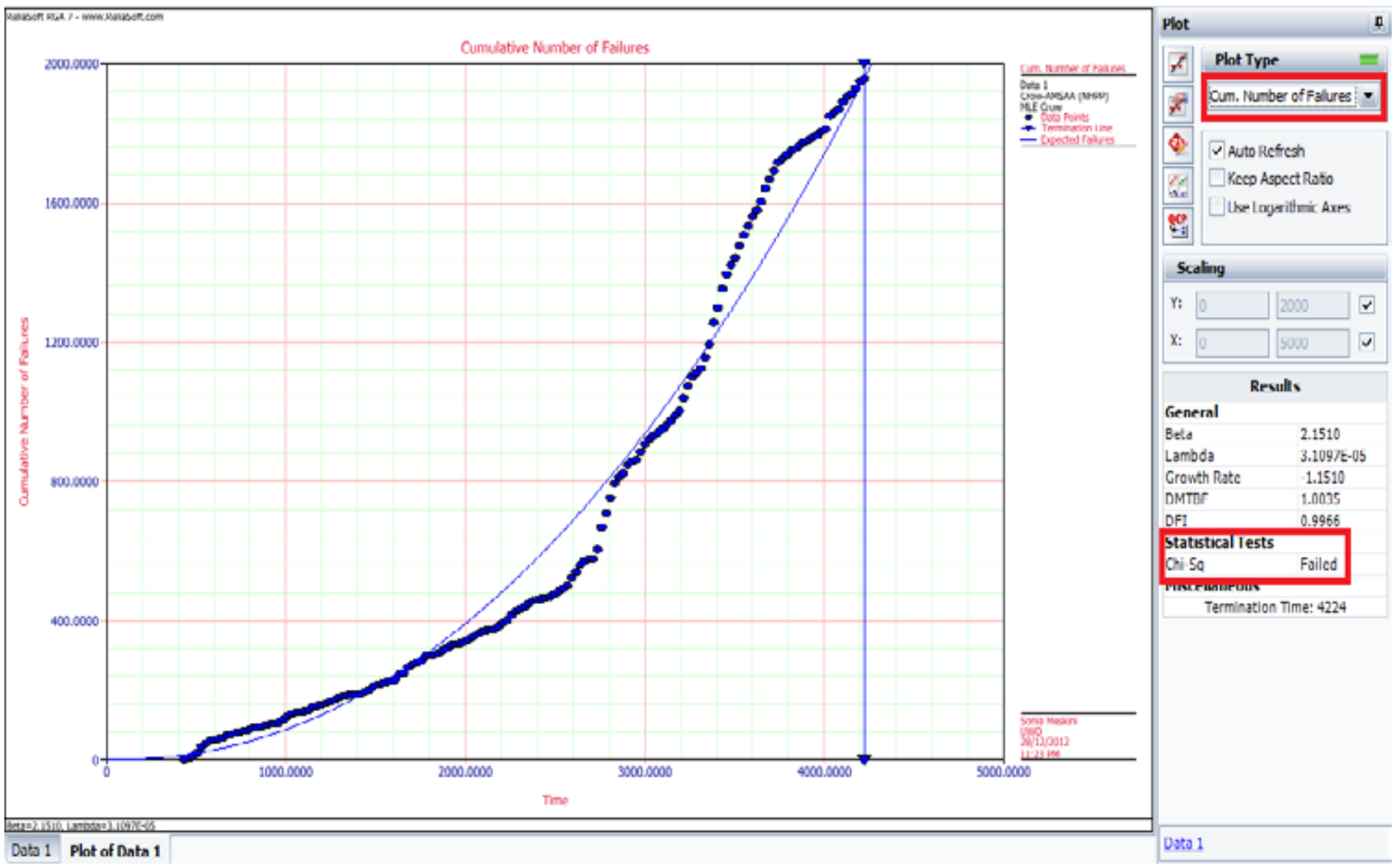

**Fig. 9.** Cumulative number of failures per time (Windows Phone data).

One explanation of the failures of the models is the goodness-of-fit tests that failed for each model application. Two tests were used: (1) the Cramer-Von Mises (CVM) test and

(2) the Chi-Square test. For both tests, the statistic (test value) has to be less than the critical value in order to have a successful fit, otherwise it is a failure, as was the case in our experiments (this was confirmed in Figure 4 by performing the CVM test, where the model failed because the statistic (0.3588) is greater that the critical value (0.173)). We had the same result, shown in Figure 9, when performing the Chi-Square test. The statistic (1282.0059) is higher than the critical value (180.0942), and the model did not accurately fir the data [15].

Thus, the most successful reliability models failed to fit all the data and to predict reliability in the mobile area for smartphone applications. This failure can be traced back to the main differences between the desktop area and smartphones. One of the mobile application failure characteristics is that they are application dependent, in the sense that they are dynamic and non-homogenously spread in time. Moreover, they are unpredictable; sometimes they decrease and sometimes they increase. One possible explanation is that reliability depends on how the application is used (Figure 3 previous section), where it is used, and when it is used. The usage may differ from one person to another, from one country to another, from one condition and time to another, etc.; this explains the uncertainty of usage of the application in the execution and release time because all these factors play an important role in the reliability of the application.

Another reason is that the DLC of a mobile application is short (up to 90 days) and the programmer aims to develop the application as fast as possible to satisfy the time to market constraint, which leads to skip phases from the DLC. The phase most often skipped is the design phase, which is the most important phase in the DLC of the application [10]. Thus, it would be difficult to identify the causes of errors during the execution time and to find a convenient solution to fix them. Besides that, the failure or unreliability of the application may be caused by the technology used during the development process. The skills of the developer and the tester also play a huge role in the reliability of the application.

Moreover, the device itself and its hardware characteristics – such as the size of the screen, the performance, the keyboard, etc. – can have a direct effect on the reliability of the application [9]. For example to adjust the map size to a certain zoom level, a zoom in/out function is needed. However, to ensure a perfect use of this function, the performance of the device must be taken into consideration [9]. Based on different surveys and studies, reliability was identified as one of the most important quality attributes of the application software [6]. Thus, the reliability of smartphone applications needs to be ensured because individuals are using their own smartphones for daily life activities and tasks more than PCs. Our study confirms that a reliability growth model adapted to smartphone applications is needed, since the traditional reliability models turned out to be inefficient.

## 4. Failure Data Analysis Using Model Distributions

The preceding section was devoted to the application of the three most used SRGMs to two common smartphone applications, Skype and Vtok, and one private Windows phone application. The inputs to these models were the instantaneous failure data, i.e, the failure number and its exact time of occurrence. Those models failed to describe adequately the failure data. One possible reason for this is that on a real time scale, the failure data of smartphone applications fluctuate highly. Having tried several non-linear models to better fit the failure data, we found that Weibull and Gamma distributions can be used to model new collected failure data of the same application after sorting them by version number and grouping them in different time periods [16]. Therefore, we used the two mentioned distributions and their particular cases, the Rayleigh and the S-Shaped models, and compared their performances for each application. This study was carried out in two steps: (1) the failure data for each application were sorted by version number and (2) the data were grouped by larger time scales (days, weeks, and months). An estimation of the total number of defects in each smartphone application version was obtained.

As many users from different regions have responded to our call for failure data collection of smartphone applications, we have received new enriched data for Skype and Vtok applications. The data were collected synchronously and come from different versions of the applications. When we plotted the raw data on the real timescale, the obtained curves were highly fluctuating and no regularity could be detected. But after sorting the data by version number and grouping the data using a larger time scale (days, weeks, months), the relationship between failure counts and time (days, weeks, months) was represented by a non-linear graph. Each application version shows an early "burst of failures" followed by a decrease where the failures became less and less frequent. For the Windows phone application, we saw in the preceding section that the failure count curve was highly fluctuating (Figure 3), but when plotted with larger time periods, it also presented the Weibull shape.

Based on this observation, we conducted a thorough study of the collected data [16]. Each version of each application was studied separately, when we had sufficient data for it. The versions with very few failure data were not considered, and they are evidently the most stable. For the Skype application we collected enough data for three versions whereas for Vtok we collected enough data for two versions. For the Windows phone application, we just grouped the failure data in larger time scales and modeled the failure count curves.

The Weibull and the Gamma distributions were characterized by two model parameters. Both contained the exponential distribution as a particular case. Excepting this case, their general shape went through a maximum and then continuously decreased. The Rayleigh and S-Shaped models, very often used in many reliability investigations, are particular cases of the Weibull and Gamma distributions, respectively. They were also used in the modeling of the observed failure data, grouped as previously indicated. A comparison between the models was carried out in each case based on the error criteria.

The Weibull distribution [24] is a two parameter function whose expression is given by:

$$f(t) = \text{wblpdf}\left(t, a, b\right) = \frac{b}{a} * \left(\frac{t}{a}\right)^{b-1} \exp\left(-\left(\frac{t}{a}\right)^{b}\right) \quad . \tag{4}$$

The parameters *a* and *b* take positive values as well as the variable *t*. If we define $A = 1/a^b$ and $B = b$, the expression simplifies to:

$$f\left(t\right) = B\ A\ t^{B-1} \exp\left(-A\ t^{B}\right). \tag{5}$$

A maximum for this function occurs at time $t = T_{max}$, such that

$$T_{max}^{b} = \frac{B-1}{A\ B}. \tag{6}$$

The Gamma distribution is a two parameter function whose expression is given by:

$$f\left(t\right) = \text{gampdf}\left(t, a, b\right) = \frac{1}{b^{a}\Gamma\left(a\right)}\left(t\right)^{a-1} \exp\left(-\frac{t}{b}\right). \tag{7}$$

for *a*, *b* and *t* taking positive values.
The maximum of this function occurs at $t = T_{max}$, such that:

$$T_{max} = b\left(a-1\right). \tag{8}$$

### 4.1. *Results*

This section presents a comparison and an evaluation of the use of the above mentioned distributions to model the failure fata of three versions of Skype (V1, V2, V3), two versions of Vtok (V1, V2), and the Windows phone application, based on the usual evaluation criteria: RMSE, Ad-R-Square, and MRE. V1, V2, and V3 are major versions, which mean that we collected the sub-versions 1.x in one major version 1. Due to space limitation, only Skype V1 and Vtok V2 will be presented from the versions we studied as well as the Windows phone application. The full and detailed results are found in [7].

For each application, the four distributions used were compared on the basis of their Root-Mean-Squared-Error (RMSE) and their Adjusted R-Square. The results of the estimated total number of defects were evaluated using the Magnitude of Relative Error (MRE).

- Root Mean Squared Error: the RMSE is the square root of the mean of the square of the differences between the actual and the predicted values and is expressed:

$$\text{RMSE} = \sqrt{\sum_{i=1}^{N} \frac{\left(y_i - \hat{y}_i\right)^2}{N}}\ . \tag{9}$$

where $y_i$ and $\hat{y}_i$ are the actual and predicted number of failures respectively, $N$ is the number of observations.

- The Magnitude of Relative Error (MRE) for each observation *i* can be obtained as follows:

$$\text{MRE} = \frac{\left| \text{C} - \hat{\text{C}} \right|}{\text{C}}. \tag{10}$$

where $C$ and $\hat{C}$ are the actual and predicted cumulative number of failures respectively.

- The Adjusted R-Square: measures the proportion of the variation in the dependent variable accounted for by the explanatory variables. Unlike R-Square, Adjusted R-Square allowed for the degrees of freedom associated with the sums of the squares. Therefore, even though the residual sum of squares decreased or remained the same as new explanatory variables were added, the residual variance did not. For this reason, Adjusted R-Square is generally considered to be a more accurate goodness-of-fit measure than R-Square. The Adjusted R-Square can be negative.

Tables 1 to 4 give a compilation of all model parameters *(a, b)* along with the predicted or estimated $T_{max}$ (time of maximum failure rate) and the expected proportion $(Y(t = T_{max})/C)$ of encountered failures by $T_{max}$. It also gives results of RMSE, As-R-Square, the estimated cumulative number of failures *C*, and the MRE shown by each model. Only the best and the second best model distributions are given for each application version. Likewise, Figure 10 to Figure 13 display a graphical comparison of all model distributions for the selected application version.

Table 1. Skype Version 1 – Error evaluation and model comparison.

| Skype V1 | Weibull | Gamma |
|---|---|---|
| Model parameters and deduced estimated values | a = 6.17 (5.26, 7.09)<br>b = 2.82 (1.81, 3.84)<br>Observed $T_{max}$ = 6<br>Estimated $T_{max}$ = 5.98<br>Estimated $(Y(t = T_{max})/C)$ = 47% | a = 6.14 (1.84, 10.44)<br>b = 0.97 (0.21, 1.73)<br>Observed $T_{max}$ = 6<br>Estimated $T_{max}$ = 5.01<br>Estimated $(Y(t = T_{max})/C)$ = 38.5% |
| RMSE | 2.1966 | 2.2305 |
| Ad-R-Square | 0.6374 | 0.6262 |
| C: Estimated cumulative number of failures or defects | 50.54 (34.51, 66.58) | 51.81 (34.32, 69.31) |
| MRE(%) | 6.4 | 4 |

Note that the value of each parameter is given in a 95% confidence interval – particularly the estimated total number of failures *C*, which has a lower and an upper bound. Only the optimum value is used to calculate the MRE. For example, for Skype Version 1, 54 failures are collected. It can be conducted from Table 1 that Weibull (with parameters: a = 6.179 and b = 2.826) is the best distribution that models the failure data, compared to the other distributions. It has the lowest RMSE: 2.1966 associated with the highest Ad-R-Square: 0.6374. The lowest MRE belongs to the Gamma distribution (with parameters a = 6.141 and b = 0.975) with a value of 4% but based on the other evaluation criteria it is the second best distribution that fitted the data.

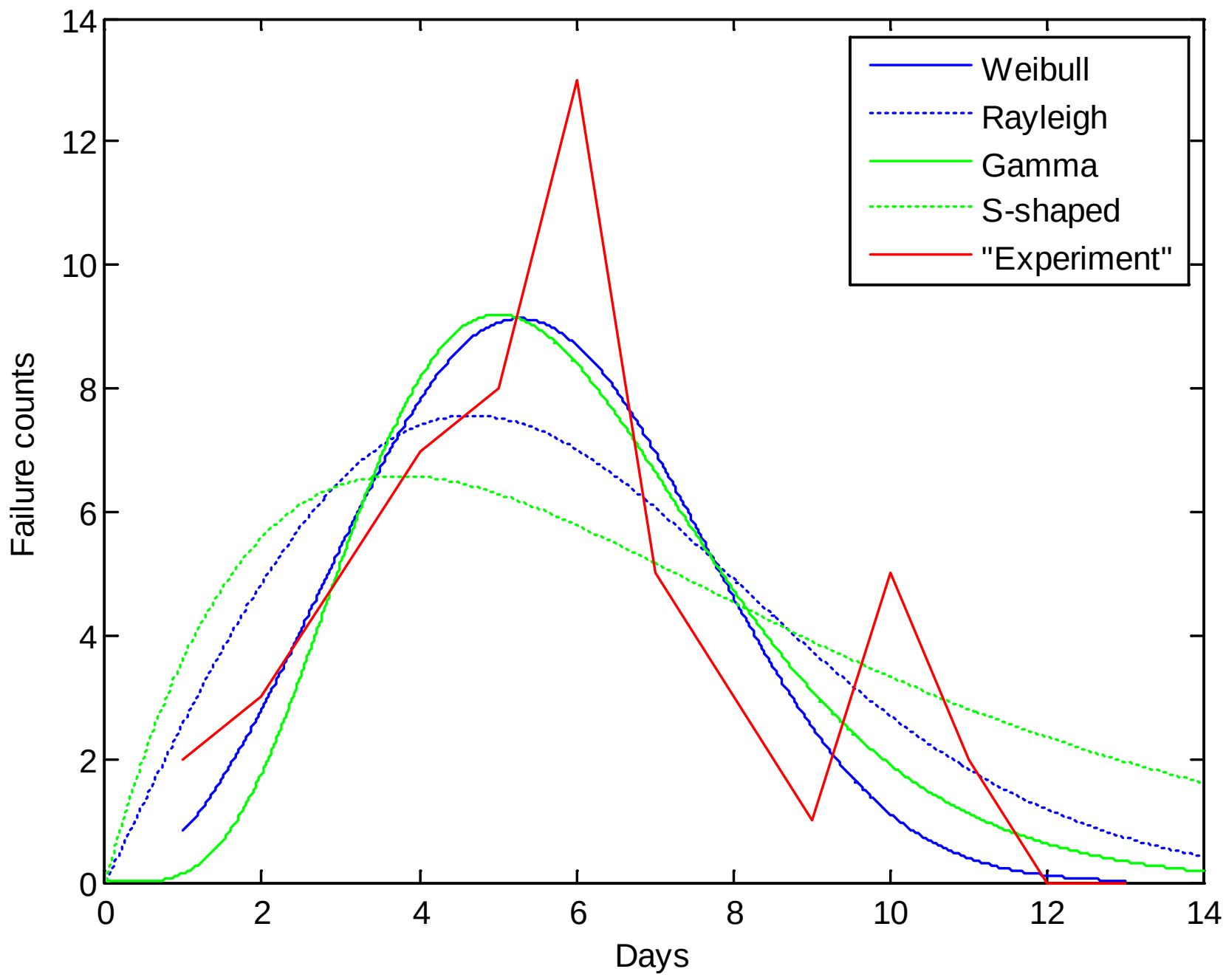

**Fig. 10.** Skype Version 1 – Model comparison.

Figure 10 portrays the results reported in Table 1. It can be noted from this figure that the Weibull distribution is the closest to the actual behavior curve of the application followed by the Gamma distribution. The difference between the actual curve and the modeling curves is explained by (1) the random nature of the failure event, which gives the spiky feature of the observed data, and (2) the size effect. Had we collected more data, the real failure curves would be smoother, but we still expect the general shapes to be explained by the chosen distributions because they capture the main behavior of the failure data of each smartphone application version. Besides that, the eventual modifications made to the application, its usage that differs from one environment to another and user to user [12] may play an important role in the behavior of the collected failure data.

Table 2. Vtok Version 2 – Error evaluation and model comparison.

| Vtok V2 | Weibull | Gamma |
|---|---|---|
| Model parameters and deduced estimated values | a = 5.75 (4.63, 6.86)<br>b = 1.79 (1.32, 2.26)<br>Observed $T_{max}$ = 3<br>Estimated $T_{max}$ = 3.64<br>Estimated $(Y(t = T_{max})/C)$ = 39% | a = 2.69 (1.55, 3.83)<br>b = 1.99 (0.89, 3.09)<br>Observed $T_{max}$ = 3<br>Estimated $T_{max}$ = 3.38<br>Estimated $(Y(t = T_{max})/C)$ = 31% |
| RMSE | 2.3401 | 2.2120 |
| Ad-R-Square | 0.7386 | 0.7665 |
| C: Estimated cumulative number of failures or defects | 80.28 (60.58, 99.99) | 80.93 (61.63, 100.2) |
| MRE(%) | 0.3 | 1.1 |

For Vtok Version 2, 80 failures were collected in 13 weeks. Based on Table 2, it can be concluded that Gamma (with parameters a = 2.696 and b = 1.996) is the best distribution that models the failure data of this application version. It has the lowest RMSE: 2.2120 and the highest Ad-R-Square: 0.7665. The lowest MRE belongs to the Weibull distribution (with parameters: a = 5.75 and b = 1.793) with a value of 0.3%. However, based on the other evaluation criteria it is the second best distribution that models the data with an Ad-R-Square value of 0.7386 and an RMSE of 2.3401. According to the Gamma model distribution, the estimated total number of defects in the worst scenario is 100.

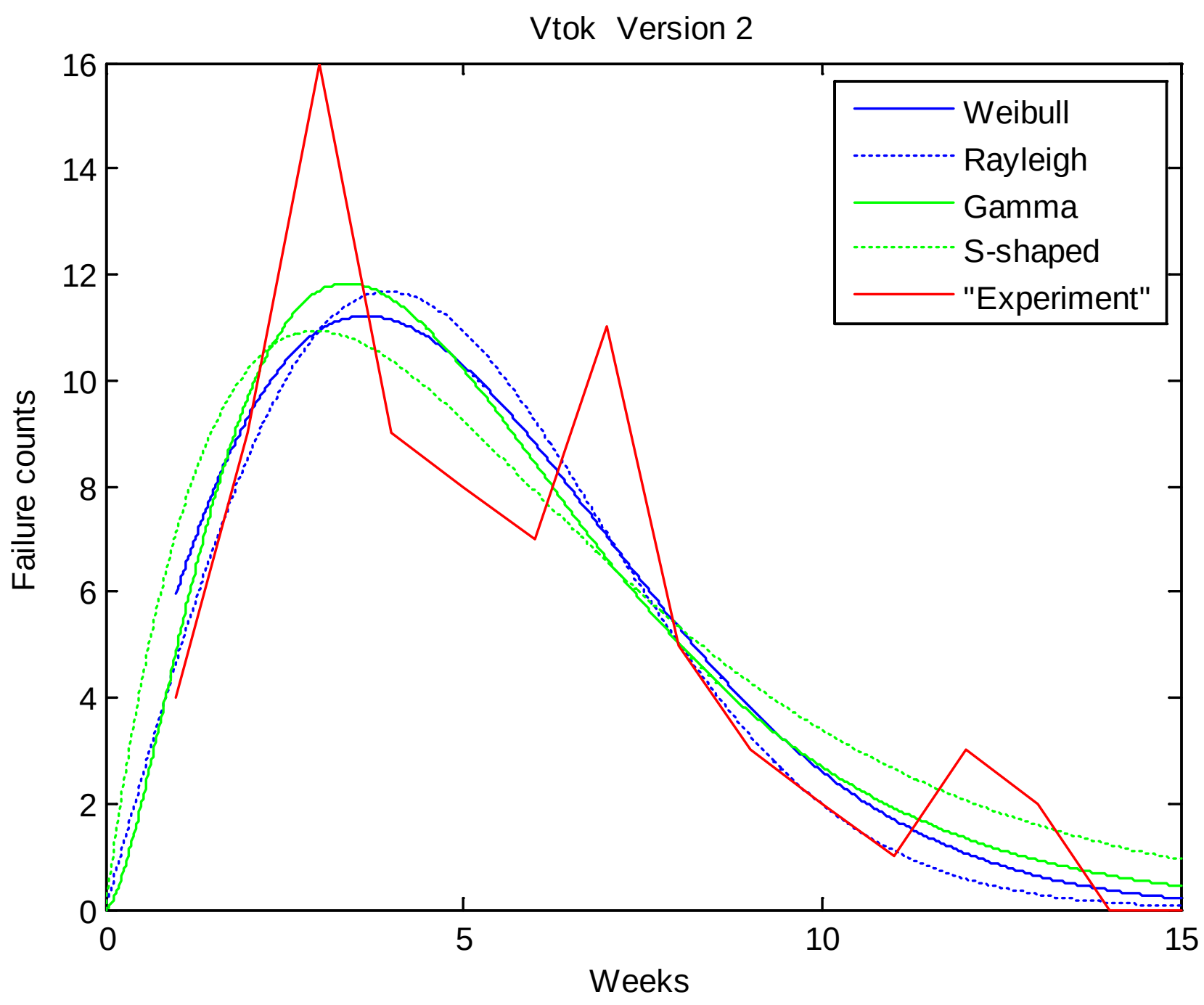

**Fig. 11.** Vtok Version 2 – Model comparison.

Figure 11 illustrates the results summarized in Table 2. It can be concluded that the Gamma distribution models adequately the failure data. It is to be noted that a small “burst” of failures occurred in the seventh week followed by a steady decrease.

As mentioned earlier, when plotted on the real time scale, the failure data of the Windows phone application are highly fluctuating and the time-to-event models could not adequately describe the data. Two time scales were adopted: the failure data are first grouped by week and in a second step are grouped by month. In each case, the same modeling carried previously was performed.

For the Windows phone application, it can be concluded that the Weibull distribution (with parameters: a = 22.11 and b = 6.248) performs better than Gamma (with parameters: a = 29.64 and b = 0724) but they do not differ markedly in modeling the failure data.

Table 3. Windows phone application (per weeks) – Error evaluation and model comparison.

| Windows phone application | Weibull | Gamma |
|---|---|---|
| Model parameters and deduced estimated values | a = 22.11 (21.09, 23.14)<br>b = 6.24 (4.47, 8.02)<br>Observed $T_{max}$ = 22<br>Estimated $T_{max}$ = 21.5<br>Estimated $(Y(t = T_{max})/C)$ = 56.83% | a = 29.64 (10.49, 48.7)<br>b = 0.72 (0.24, 1.20)<br>Observed $T_{max}$ = 22<br>Estimated $T_{max}$ = 20.75<br>Estimated $(Y(t = T_{max})/C)$ = 45% |
| RMSE | 44.2987 | 47.5420 |
| Ad-R-Square | 0.6112 | 0.5521 |
| C: Estimated cumulative number of failures or defects | 1722 (1315, 2130) | 1721 (1245, 2198) |
| MRE(%) | 12 | 12.05 |

Table 4. Windows phone application (per months) – Error evaluation and model comparison.

| Windows phone application | Weibull | Gamma |
|---|---|---|
| Model parameters and deduced estimated values | a = 5.20 (4.94, 5.46)<br>b = 5.42 (3.95, 6.88)<br>Observed $T_{max}$ = 5<br>Estimated $T_{max}$ = 5.01<br>Estimated $(Y(t = T_{max})/C)$ = 55.76% | a = 25.48 (2.74, 48.21)<br>b = 0.19 (0.01, 0.37)<br>Observed $T_{max}$ = 5<br>Estimated $T_{max}$ = 4.82<br>Estimated $(Y(t = T_{max})/C)$ = 44.65% |
| RMSE | 64.7462 | 106.89 |
| Ad-R-Square | 0.9363 | 0.8264 |
| C: Estimated cumulative number of failures or defects | 1831 (1420, 2243) | 1767 (1081, 2454) |
| MRE(%) | 6.43 | 9.7 |

Figure 12 depicts the results summarized in Table 3. It is to be noted that, even portioned in weeks, the Windows phone application failure data are still not smooth but the general trend is reproduced by the two models.

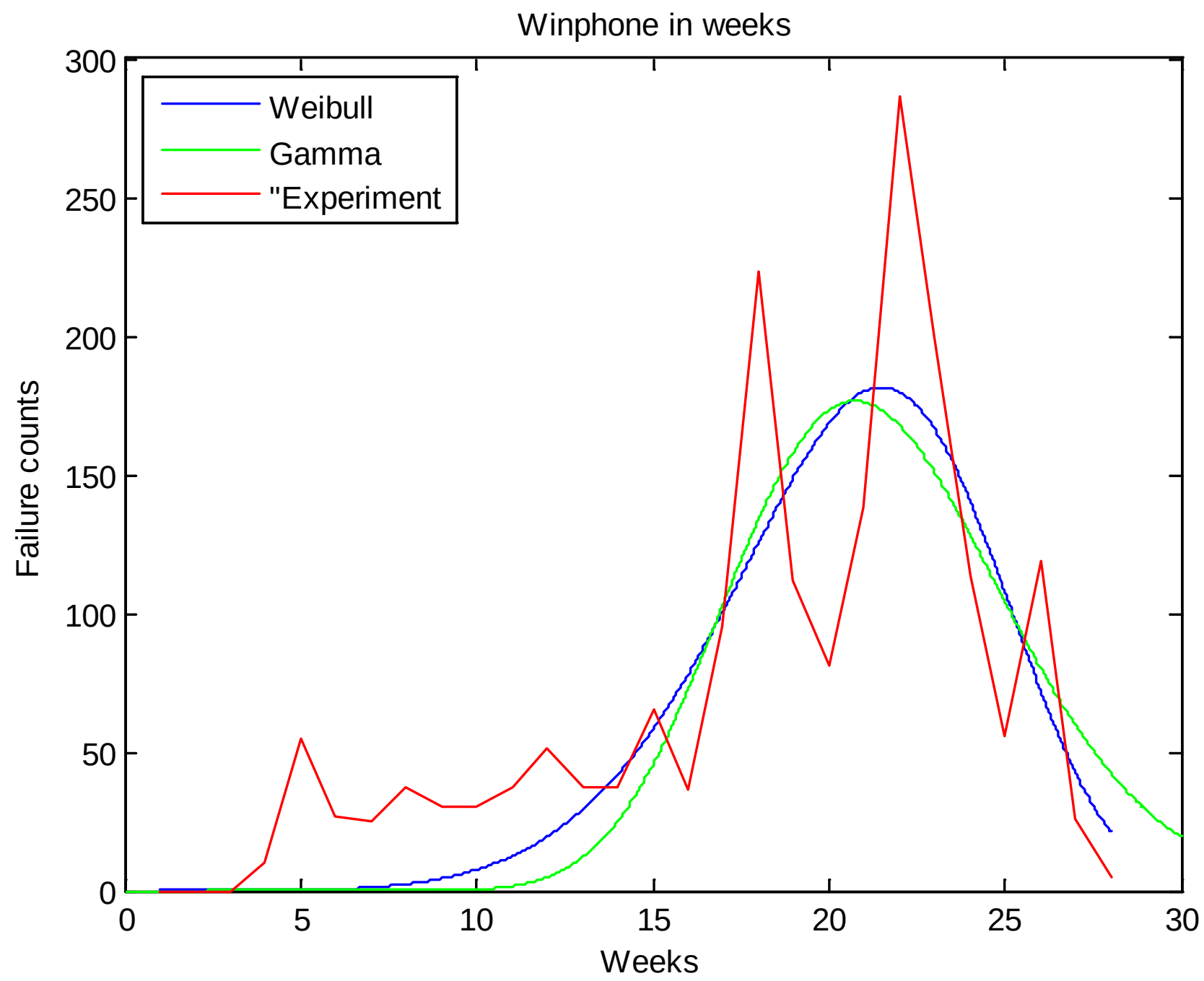

**Fig. 12.** Windows phone application failure data per week – Model comparison.

A further grouping of the Windows phone application failure data by months and the above analysis was carried out. Based on Table 4, it can be concluded that, again, the Weibull (with parameters: a = 5.20 and b = 5.42) model distribution performs better than Gamma (with parameters: a = 25.48 and b = 0.196).

Figure 13 illustrates the results summarized in Table 4. It can be concluded that the Weibull distribution adequately models the failure data.

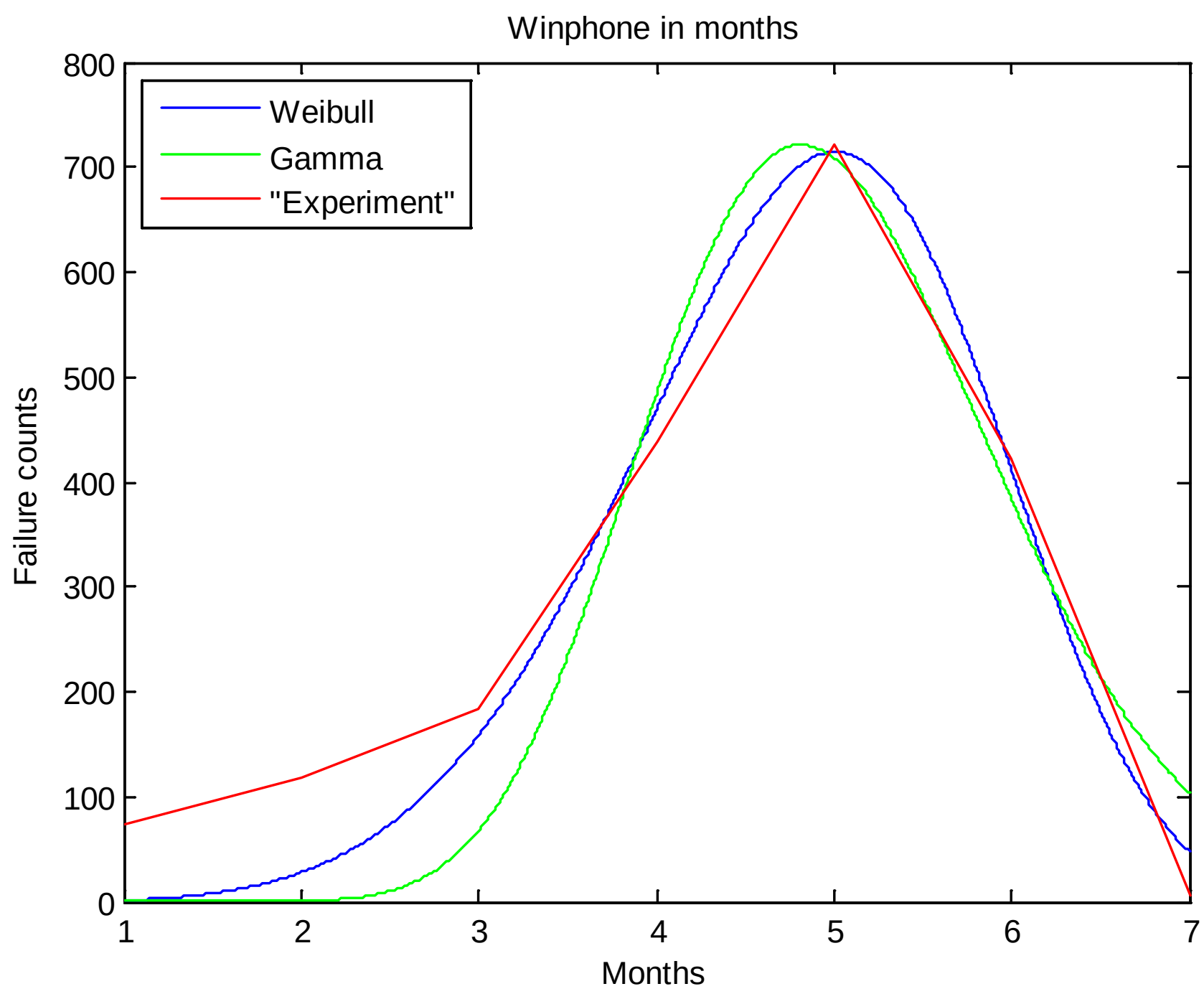

**Fig. 13.** Windows phone application failure data per month – Model comparison.

## 5. Discussion

According to the preceding section, and as an answer to the first research question raised in the abstract, it can be concluded that the most successful reliability models [17, 18] failed to account for the failure data and to predict the reliability of mobile applications. This failure can be traced back to the following main reasons:

### 5.1. *Operational Environment and Usage Profiles of Smartphones Applications*

One of the mobile application failure characteristics is that they are application dependent in the sense that they are dynamic and non-homogenously spread in time [19]. Moreover, they are unpredictable, happening in "bursts". One possible explanation is that reliability depends on the type of the application used (example the Windows phone application mentioned earlier), and on its operational environment and usage profile (where and when and how the application is used). The usage may differ from one user to another, from one country to another, from one condition to another, etc. [12], which explains the uncertainty of usage of the application in the execution and release time [10]; all of these factors play an important role in the reliability of the applications.

Another reason is that the DLC of a mobile application is short (up to 90 days) and the programmer aims to develop the application as fast as possible to satisfy the time to market constraint; this can lead to skipping phases from the DLC. The phase most often skipped is the design phase, which is the most important phase in the DLC of the application [3]. Thus, it is difficult to identify the causes of errors during the execution time and to find a convenient solution to fix them. Besides that, the failure or unreliability of the application may be caused by the technology used during the development process. Also, the skills of the developer and the tester play a huge role in the reliability of the application

### 5.2. *Hardware and Software Limitations*

The device itself and its hardware characteristics – such as the size of the screen, the performance, the keyboard, etc. – can have a direct effect on the reliability of the application. For example, to adjust the map size to a certain zoom level, a zoom in/out function is needed. However, to ensure a perfect use of this function the performance of the device has to be taken into consideration [10].

Other reasons that may explain the dynamic aspect of smartphone applications and the various possible causes of their observed crashes are detailed in [20], which gives an idea about the different causes, external and internal, of the unreliability of the application.

### 5.3. *Need to Re-examine the Standard Assumptions*

Software managers usually base their assessment of the reliability of software and its future evolution (after release) on a simple extrapolation in time of a reliability model, based on the failure data collected in the testing phase [21]. They implicitly assume that the field operational use of the application will not differ greatly from that of the lab testing phase [22, 23]. This is the first standard assumption on which most of the reliability models are based.

This assumption is no longer valid in the mobile area as there are large variations and uncertainties in the operational profile of smartphone applications. There are as many possible operational profiles, as there are millions of users. If an application is used in different environments, its reliability may be different for each environment [12]. Therefore, the mobile feature has to be included in the initial assumptions of any reliability model suited to the mobile area in order for its assessments and predictions to be taken seriously.

### 5.4. *The Answers to the Research Questions*

Assuming all of these uncertainties, at a second stage, and in order to answer the second research question, we collected data from many users in different regions of the world, sorted them by application versions, and grouped them in different time periods (days, weeks, and months). Each application version failure data, when plotted in time periods, shows the same pattern: an early “burst of failures”, due probably to the most evident defects, followed by a step decrease in failure rate. After trying several non-linear models to fit the failure data, we found that the observed behavior is better modeled by the Weibull or Gamma distributions. To answer the third research question the main features of this approach can be summarized as follows:

- For each application version, the model distributions are in fact distinguished by tiny differences in the calculated errors RMSE and Ad-R-Squared. Nevertheless, it can be concluded that no one single distribution can fit the data of all applications or even the different versions of the same application.
- As the parameters are given along with their 95% confidence intervals, it is to be noted that parameter $b$ of the Weibull distribution, which is fixed to the value $b = 2$ for the particular case of Rayleigh distribution, has confidence intervals that include the value $b = 2$. The same can be noted for parameter $a$ of the Gamma distribution, which is fixed to the value $a = 2$ in the particular case of the S-shaped model distribution. But most of the time, the general distribution models fit the failure data better than the particular cases.
- Similar to the famous 40% rule of the Rayleigh distribution, and independent of any application, the S-shaped distribution has a 26.4% rule. This means that by $T_{max}$, only 26.4% of the defects in a smartphone application will be uncovered. This can be tested on larger datasets and across many applications.
- For all the applications, the estimated and observed $T_{max}$ are in good agreement. For the particular case of the Windows phone application only the Weibull and Gamma distributions can model the data because the value of parameter $b$ for Weibull and that of parameter $a$ for Gamma are greater than 2.
- It can be noted that the Gamma distribution, along with its particular S-shaped case, model more frequently the observed failure data.

## 6. Threats to Validity

As studying the reliability of smartphone applications is a new approach and no previous work or techniques were suggested, some limitations can become a threat to the validation of our approach, such as:

- The data collection was a challenge and insufficient data were collected due to the privacy policies of smartphone application developers and users who believed that someone could control their phone from the crash report of their applications. Therefore, the limited datasets of our investigations make a generalization of the results difficult.
- Experimentation with more elaborate data and other model distributions could be investigated. On the other hand, data from different platforms, like Android and its users, will be useful to get more accurate results.
- Only a few versions were used in this study. A study of more versions of the applications would help get a better idea whether the version failure data are independent of each other or correlated in a particular way.

- It is not guaranteed that the modifications made to the application from one version to another would not introduce more bugs or complexity. Moreover, it cannot be promised that the correction of old bugs would not result in other bugs.
- The data were collected during the operational phases of the applications. The early development stages of each application and its architecture, whether simple or complex, are unknown. Besides that, as mentioned earlier, the operational conditions and usage profiles are also unknown. All of these factors may create defects than expected.

## 7. Conclusions

Our work is a step toward the application and evaluation of traditional Software Reliability models in the mobile area. We selected three of the most used models that are known for their efficiency in the desktop area: the NHPP, Musa-Basic, and Musa-Okumoto models. We examined two iPhone applications, Skype and Vtok, which were used and tested differently to evaluate the models under different conditions, and one Windows phone application that whose name we didn't mention because of the company's confidentiality policies. It turned out that none of the selected SRGMs was able to account for the failure data satisfactorily.

Our study also highlighted the causes of the failure of the models and the need for a meticulous SRGM for Smartphone applications, because the existing software reliability approaches were developed for traditional desktop software applications that are static and stable during their execution. This is not the case for smartphone applications, which have an unknown operational profile, a highly dynamic configuration, and changing execution conditions. On a continuous background, the smartphone failures come in relatively short bursts from time to time, which explain the abrupt in the observed cumulative failure number curves. This particular feature cannot be accommodated by the SRGMs that were used. Thus, in order to evaluate the reliability of smartphone applications, new models, principles, and tools are needed to incorporate the underlying uncertainties of such applications [19].

Our investigation of smartphone application reliability through the use of well-known available growth models suited primarily to desktop applications is twofold: (1) highlight the versatile nature of mobile applications, their dynamic configuration, unknown operational profile, and varying execution conditions in contrast to the static and stable desktop ones, and (2) stress the need for the design of new reliability models suited for mobile applications that take into account the inherent versatility of such applications [19].

As is well known, reliability is one of the most important features of an application and great efforts have been devoted to tailor and predict it through the study of recorded failure data. A non-reliable application means dissatisfied customers, loss of market share, and significant costs to the supplier. For critical applications, such as banking or health monitoring, non-reliability can lead to great damage. Therefore, it is of great importance to ensure early detection and resolution of reliability issues in desktop applications as well as, increasingly, in mobile applications.

Our future work will focus on analyzing these selected SRGMs in more depth and trying to modify the closest one to the data and adapt in to smartphone applications. Moreover, we will check to find out if we need to have a specific model for each type of applications or if one model is applicable to all the categories, of Smartphone applications, taking into consideration the severity of the failure.

Another future project would be to evaluate the possibility of applying more than one model on the same failure data, such as the Windows phone crash count, by dividing the data into two or more categories and applying the convenient model to each category to predict the reliability of the application. Further investigations of Android failure data are also underway.

## Conflict of Interest

The authors declare that they have no conflict interest (financial or non-financial) related to this research.

## Funding

This study has been funded by the National Science and Engineering Research Council (NSERC) of Canada, grant number RGPIN-2020-04325.